\documentclass[twocolumn]{aa} 

\newcommand{\htwo}{H$_2$}
\newcommand{\htwoo}{H$_2$O}
\newcommand{\otwo}{O$_2$}

\newcommand{\hp}{H$^+$}

\newcommand{\hthreeop}{H$_3$O$^+$}

\newcommand{\op}{O$^+$}
\newcommand{\opp}{O$^{2+}$}

\newcommand{\otwop}{O$_2^+$}

\newcommand{\hi} {H {\sc i}}

\newcommand{\cii}{C {\sc ii}}

\newcommand{\pimenc}{$\pi$ Men c}
\newcommand{\trappistoneb}{TRAPPIST-1 b}

\newcommand{\othreep}{O($^3$$P$)}
\newcommand{\ooned}{O($^1$$D$)}
\newcommand{\oones}{O($^1$$S$)}

\newcommand{\opfours}{O$^+$($^4$$S^{\rm{o}}$)}
\newcommand{\optwod}{O$^+$($^2$$D^{\rm{o}}$)}
\newcommand{\optwop}{O$^+$($^2$$P^{\rm{o}}$)}

\newcommand{\lalpha}{Lyman-$\alpha$}

\usepackage{amsmath}
\usepackage{mathtools}
\usepackage{ulem}

\usepackage{placeins}

\usepackage{graphicx}
\usepackage{txfonts}
\usepackage{xcolor}
\usepackage{natbib}
\begin{document} 

\title{Heating and ionization by non-thermal electrons in the upper atmospheres of water-rich exoplanets}

   \author{A. Garc\'ia Mu\~noz
          \inst{1}
          }

\institute{
Universit\'e Paris-Saclay, Universit\'e Paris Cit\'e, CEA, CNRS, AIM, 91191, Gif-sur-Yvette, France\\
\email{antonio.garciamunoz@cea.fr}}


 
  \abstract 
  {The long-term evolution of an atmosphere and the remote detectability of its chemical constituents are susceptible to how the atmospheric gas responds to stellar irradiation. 
  The response remains poorly characterized for water and its dissociation products, however, this knowledge is relevant to  our understanding of hypothetical water-rich exoplanets.
  }
  {Our work investigates the effect of photoelectrons, namely,
  the non-thermal electrons produced by photoionizing stellar radiation on the heating and ionization of extended atmospheres dominated by the dissociation products of water.
  }
  {We used a Monte Carlo model and up-to-date collision cross sections  
  to simulate the slowing down of photoelectrons 
  in O-H mixtures for a range of fractional ionizations and photoelectron energies.\footnote{
  The Monte Carlo model is available through https://antoniogarciamunoz.wordpress.com/ or by e-mail request at antonio.garciamunoz@cea.fr or tonhingm@gmail.com. The package includes the FORTRAN sources, the implemented cross sections and a variety of input/output files that include the data needed to reproduce Figs. \ref{eta_yield_steam_fig}-\ref{eta_inelas_steam_fig} of the paper.  
  }
  }
  {We find that that the fraction of energy of a photoelectron 
  that goes into heating is  similar in a pure H gas and in O-H mixtures, except for very low fractional ionizations, whereby the O atom remains an efficient sink of energy.
 The O-H mixtures will  go on to produce more electrons because the O atom is particularly susceptible to ionization. We quantified all that information and present it
 in a way that can be easily incorporated into photochemical-hydrodynamical models.
  }
  {Neglecting the role of photoelectrons in models of water-rich atmospheres will result in overestimations of the atmospheric heating and, foreseeably, the mass-loss rates as well. 
  It will also  underestimate the rate at which the atmospheric gas becomes ionized, which may have implications for
  the detection of extended atmospheres with {\lalpha} transmission spectroscopy. 
  Our simulations for the small exoplanets {\pimenc} and {\trappistoneb} reveal that they respond very differently to irradiation from their host stars, with water remaining in molecular form  at lower pressures in the latter case.
  }


   \maketitle
%

\section{Introduction}

The fact that planets with water-rich atmospheres may be common in our galaxy is a remarkable possibility, as such planets are not known to occur in our vicinity.
With varying degrees of uncertainty, this idea is supported by the following points: 
{\textit{i)} planetary formation and evolution theories predicting that significant water is accreted during the embryo formation if it takes place beyond the snowline -- or at a later time and possibly closer to the host star. This process would occur while the planet is being bombarded by icy planetesimals or from chemical reaction between a thick primordial envelope and an oxidizing magma ocean \citep{venturinietal2020,bitschetal2021,kimuraikoma2022};
\textit{ii)} 
the constraints inferred from  interior structure models 
on the water mass fraction for individual planets with measured mass, $M_{\rm{p}}$, and radius, $R_{\rm{p}}$ \citep{valenciaetal2010,delrezetal2021}; 
\textit{iii)} 
a characteristic dispersion in the $M_{\rm{p}}$-$R_{\rm{p}}$ data, suggesting a  transition between rocky planets with essentially no  envelopes
for $M_{\rm{p}}$/$M_{\oplus}${$\lesssim$}5 and planets with a broad range of compositions for $M_{\rm{p}}$/$M_{\oplus}${$\gtrsim$}5 \citep{otegietal2020}; 
\textit{iv)}
a peak in the occurrence of planets orbiting M dwarfs for bulk densities 
consistent with high water abundances and possibly inconsistent with planets that lack an atmosphere or that have one dominated by H$_2$/He \citep{luquepalle2022}.
}

The state of the water in the planets also stands as an open debate. Steam has been identified with transmission and emission spectroscopy in a number of exoplanets  \citep{evansetal2017,bennekeetal2019,tsiarasetal2019}. In addition, water might occur in the solid, liquid, and plasma phases, or in a supercritical state
\citep{nettelmannetal2010,madhusudhanetal2020,mousisetal2020}. 
 Significant attention will be devoted to water as a key  tracer
 in exoplanet atmospheres in the coming years, notably because the JWST and ARIEL 
\citep{tinettietal2018}
telescopes have the capacity to probe the molecule in the infrared.
The upcoming observations 
might firmly establish the existence and general properties of water-rich atmospheres.

Further insights into the water content of a strongly irradiated atmosphere can be gained by investigating its long-term stability. 
The mass in the atmosphere of a small planet of known $M_{\rm{p}}$ and $R_{\rm{p}}$
may be orders of magnitude larger if the gas has a high molecular weight than if it is composed of {\htwo}/He. 
This means that a high-molecular weight atmosphere is, in principle, more stable 
even when a vigorous outflow driven by stellar irradiation is established.
For example, 
 \citet{garciamunozetal2020,garciamunozetal2021} predicted that the sub-Neptune {\pimenc} 
 in its current orbit might lose its atmosphere in a few Myrs if this is composed of {\htwo}/He, which is much less than the estimated $\sim$5 Gyr age of the system, 
but that such a catastrophic fate would take longer by $\times$10$^3$  if the atmospheric composition is dominated by {\htwoo}. 
These arguments, along with supplementary  arguments based on the detection of {\cii} and the  non-detection of {\hi} {\lalpha} in the upper atmosphere of the planet, led the authors to propose that {\pimenc} has a high-molecular weight and (possibly) a water-rich atmosphere. 

As such analyses rely on model-predicted mass loss rates, it is important to assess the completeness of the models, particularly in their treatment of the  collisional-radiative processes that occur in the low-pressure layers where the outflow is established
\citep{garciamunozetal2020,garciamunozetal2021,shaikhislamovetal2020}. It is worth noting that 
the young Earth and Venus may have  had water-rich atmospheres, a possibility that has been explored to better understand their histories
 \citep{zahnleetal1988,chassefiere1996,lichteneggeretal2016}. 
It is thus no surprise that a new generation of models continue to explore
the outflow of water-rich atmospheres 
\citep{kurosakietal2014,lugerbarnes2015,guo2019,johnstone2020,garciamunozetal2020,garciamunozetal2021,yoshidaetal2022} and the relevance of related ideas to exoplanets.

We are currently revisiting the formulation of the collisional-radiative processes that occur in the upper atmospheres of exoplanets \citep{garciamunozschneider2019}.
We believe that past modeling may not have captured all the complexity
implicit in a high-temperature, multi-component plasma that transitions from optically thick to optically thin conditions. 
The ultimate goals of this effort are to obtain accurate mass loss rates for a range of atmospheric compositions and stellar hosts, 
and set up a physics-based framework for the interpretation of the spectroscopic searches that have been conducted at some exoplanets. 
In particular, the current work addresses the fundamental issue of tracking the formation and slowing down of photoelectrons in water-rich atmospheres, as this partly dictates the fraction of the stellar irradiation that turns into heating and drives the outflow at those planets. 
That information is often quantified in the form of so-called heating efficiencies and production yields.

Calculations of heating efficiencies and production yields for general astrophysical application are available, for example, for H$_2$-H mixtures and for pure O and {\otwo} 
 \citep[e.g.][and refs. therein]{greenetal1977,victoretal1994,dalgarnoetal1999,garciamunoz2023}. 
Some calculations have also been reported for exoplanets with hydrogen-dominated atmospheres \citep{cecchi-pestellinietal2006,cecchi-pestellinietal2009,shematovichetal2014,guoben-jaffel2016}. 
 We have however not found similar calculations for {\htwoo} nor O-H mixtures. 
There is a long history of simulations of the slowing down of photoelectrons in 
the {\htwoo}-dominated comae of comets, but they emphasize the excitation of neutrals and ions that can be remotely sensed rather
than the deposition of energy; furthermore, they are mostly concerned with conditions of low fractional ionization \citep{ashihara1978,bhardwaj2003}. 
For completeness, it should be mentioned that some recent 
 works related to Earth-like \citep{tianetal2008}
and water-rich atmospheres \citep{johnstone2020,nakayamaetal2022}
have treated the problem of  heating and ionization by photoelectrons; however, it is difficult to judge the accuracy of their approximate treatments and the significance of the phenomenon in the conditions of their simulations, as compared to more general ones.

The above arguments motivate us to review the problem of photoelectrons in water-rich atmospheres
 in a way that is reliable and easy to implement into atmospheric models. 
We focus here on O-H mixtures, namely, it is assumed that the {\htwoo} has dissociated, 
but will address the more general problem of atmospheres that contain 
this and other molecules in undissociated form in future work. 
 In general, this simplification is not critical, as  water dissociates rapidly at close-in exoplanets when they orbit stars that are not too cool  \citep{guo2019,johnstone2020,garciamunozetal2020,garciamunozetal2021}. 
 Even when the star is cool (and its far-ultraviolet emission weak), a calculation based on an O-H mixture provides valuable insight into what happens in the region where the XUV photons (wavelengths shorter than the Lyman continuum threshold) are deposited. 
 We intentionally avoid approximate methods \citep{petersonetal1978} to estimate the heating efficiencies and production yields
 of mixtures based on the properties of their individual constituents 
  to keep a better control of the relevant physics and because such approximations become inaccurate for the conditions of small-to-moderate photoelectron energies that are relevant here.

The paper is organized as follows. Section 
{}\ref{hi_oi_sec} offers a discussion of the heating efficiency and production yields in O-H mixtures. Section {}\ref{application_sec} explores the heating and ionization of 
the postulated water-rich atmospheres of the exoplanets {\pimenc} and {\trappistoneb}, and estimates the lifetime of {\ooned}
in them. Lastly, 
{Section }\ref{summary_sec} summarizes the main findings and lists some possible next steps for future studies.

\begin{figure*}[h]
   \centering
   \includegraphics[width=15.cm]{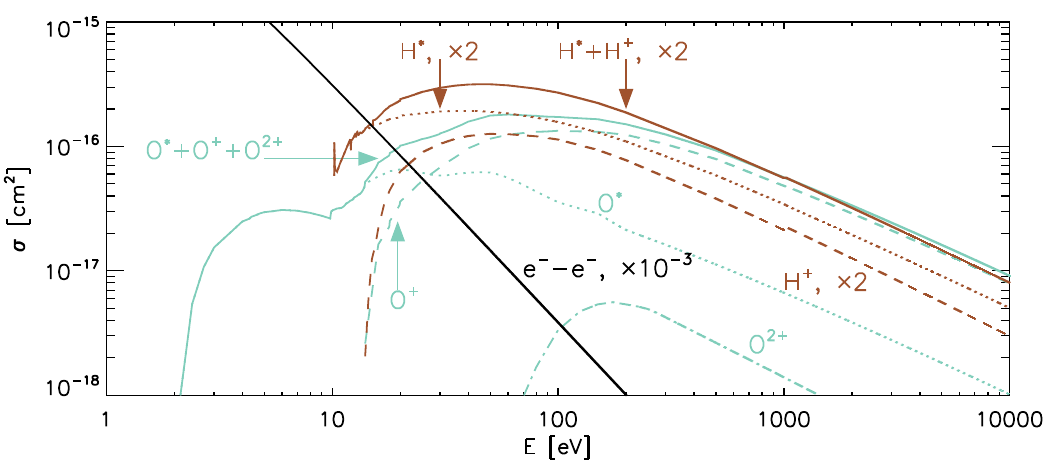}      
   \caption{\label{xs_steam_fig}
    Cross sections for 
   excitation (H$^*$, O$^*$), ionization (single, H$^+$, O$^+$, and double, {\opp}) and total inelastic (=excitation + ionization) collisions of photoelectrons with ground-level
   H and O atoms (LG90 set, see Appendix \ref{xs_appendix}). The H cross sections have been multiplied by $\times$2 to better reveal the significance of collisions with both types of atoms in an O-H mixture with $f_{\rm{O_n}/\rm{H_n}}$=1/2. 
   We also show the adopted electron-electron pseudo-cross sections multiplied by $\times$10$^{-3}$.}
   \end{figure*}

\section{\label{hi_oi_sec} Slowing down of photoelectrons in the O-H mixture}

Photoelectrons are produced in the interaction of ionizing stellar photons with the atmospheric gas,  in the process acquiring kinetic energies of up to hundreds of eV. 
The specifics of photoelectron formation are dictated by the properties of the local atmosphere and the stellar radiation that reaches it.
Once released into the atmosphere,
the nascent photoelectrons collide repeatedly with the thermal electrons and heavy particles (atoms in this work), exchanging at each collision some energy with them until the electrons are ultimately thermalized. The interactions can be elastic:
$$
 e' + {e_{\rm{th}}}, X  \xrightarrow{} e'' + {e_{\rm{th}}}, X,$$
if the exchange with the thermal electrons, 
$e_{\rm{th}}$, or the heavy particles, $X,$ involves only kinetic energies 
-- or inelastic:
 $$
 e' + X \xrightarrow{ } e'' + X^*, e'' + e''' + X^+, 
$$
if the heavy particle becomes excited or ionized
(by ejecting one electron in the example).
 Superelastic collisions, the reverse of inelastic collisions, also occur,  especially for high $X^*$ or $X^+$ densities, but their significance is often minor and easier to assess on a case-by-case basis; 
 $e'$ refers to the incident photoelectron and $e''$ and $e'''$ refer to the post-collision photoelectrons. 
 We use the term photoelectrons as equivalent to non-thermal electrons to describe the population of electrons that cannot be ascribed to the dominant single-temperature Maxwellian distribution of velocities. We note that other works use 
 the term photoelectrons to designate the primary electrons produced directly by photoionization and 
 keep the term secondary electrons to refer to the electrons that form as the primary electrons slow down and eject additional electrons in collisions with the heavy particles.
 In our use of these terms, $e'$, $e''$, and $e'''$ qualify as photoelectrons as long as their energies are sufficiently larger than the mean kinetic energy of the thermal electrons.

 In the O-H gas that we consider, $X$ may be any of the H, O, or {\op}
atoms. 
Because the electron-to-proton mass is {$\ll$}1, 
elastic collisions with {\hp} have a negligible effect compared to elastic electron-electron collisions and are omitted, and  we also omit collisions of photoelectrons with {\opp} and higher ions as their densities are expected to be low. 
In elastic collisions, the energy lost by the photoelectron is transferred to the translational energy of the thermal electrons and heavy particles, thereby heating the gas. 
In inelastic collisions, some of the photoelectron energy is expended to promote or eject a bound electron initially in the heavy particle. This expense represents in principle a sink for the thermal budget of the gas because it does not contribute to its translational energy. Some of that excitation or ionization energy may, however, eventually convert into heat if the heavy particle undergoes a non-radiative collision.

To simulate the slowing down of the photoelectrons in an O-H mixture, we used the Monte Carlo (MC) model developed in \citet{garciamunoz2023}. 
The model has been expanded 
(the original version included only H and He atoms; see Appendix \ref{xs_appendix}) 
to take into account collisions of the photoelectrons with O and {\op}. 
Our numerical experiments revealed that the role of collisions with {\op} is minor for the conditions of
moderate energy and fractional ionization of interest here, in agreement with past findings for a pure O gas \citep{kozmafransson1992,victoretal1994}, 
yet the corresponding collisional channels are kept for completeness. 
For reference, 
Fig. \ref{xs_steam_fig} shows the excitation and ionization cross sections for collisions 
with H and O. The former have been scaled by $\times$2 to clearly reveal the relative contributions in a O-H gas resulting from water dissociation.

Many of the MC model outcomes can be traced to the collisional properties of the atoms in the mixture, so it is worth noting some of them.
Figure \ref{xs_steam_fig} shows that the inelastic (excitation + ionization) cross sections of H and O are comparable in magnitude, especially at energies of {$\gtrsim$}100 eV. Therefore, in an O-H gas resulting from water dissociation, 
the photoelectrons will interact with either type of atom with comparable probabilities. 
For energies of {$\gtrsim$30 eV}, the cross section for H excitation is larger by about $\times$2 than the cross section for H ionization.
In contrast, over the same range of energies, the cross sections for O ionization are nearly an order of magnitude larger than those for O excitation. As a consequence ionization, rather than excitation, should dominate the loss of energy for the photoelectrons interacting with the O atoms.

   \begin{figure*}[t]
   \centering
   \includegraphics[width=14.cm]{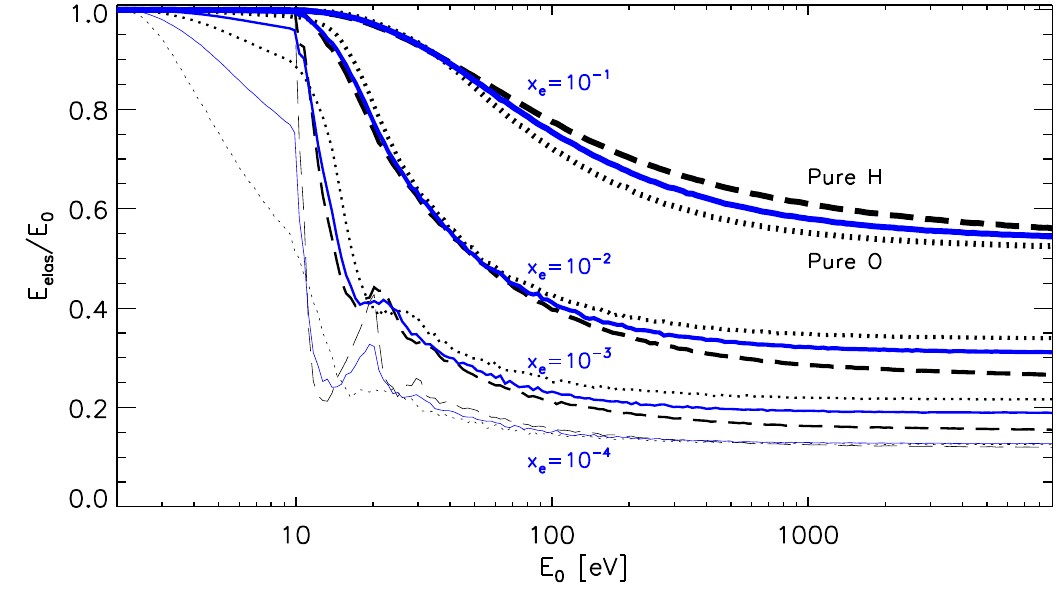}      
   \includegraphics[width=14.cm]{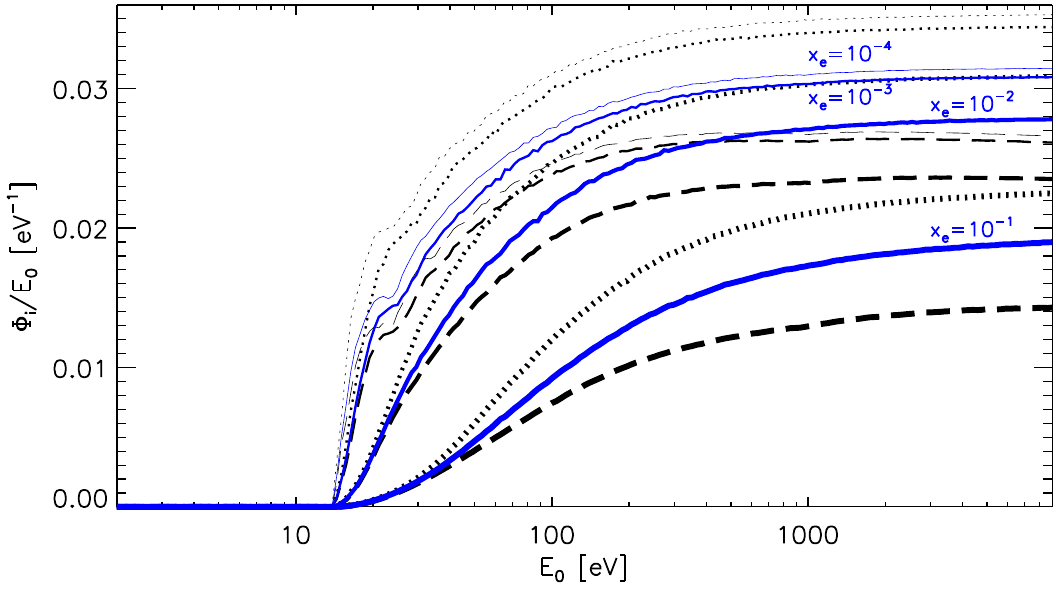}
   \caption{\label{eta_yield_steam_fig}
   Some calculated properties from the slowing down of photoelectrons.   
   Top: Fraction of the photoelectron initial energy $E_0$ that is transferred to the gas in elastic collisions. 
   Bottom: Number of secondary ions created by a photoelectron of initial energy $E_0$, normalized by $E_0$. 
   The three color-style combinations refer to 
   pure H (black, long dashed), pure O (black, dotted), and an O-H mixture with $f_{\rm{O_n}/\rm{H_n}}${=}1/2 (blue, solid). For each mixture, the four curves that are displayed (distinguishable by their thickness) refer to fractional ionizations $\log_{10}(x_{\rm{e}})$=$-1$ (thickest), $-2$, $-3,$ and $-4$ (thinnest). 
   Although not shown for the sake of clarity, the $E_{\rm{elas}}$/$E_0$ ratio for $E_0${$\gtrsim$}100 eV for pure H becomes essentially independent of fractional ionization at $x_{\rm{e}}${$<$}10$^{-4}$. In contrast, $E_{\rm{elas}}$/$E_0$ continues to drop with $x_{\rm{e}}$ for the other O-H mixtures. For example, for $E_0${$\gtrsim$}200 eV and $x_{\rm{e}}${$\sim$}10$^{-6}$, $E_{\rm{elas}}$/$E_0${$\sim$}0.05, and 0.07 for pure O and the O-H mixture with $f_{\rm{O_n}/\rm{H_n}}${=}1/2, respectively.  
   Calculations were done with the LG90 set of cross sections.
   }
   \end{figure*}

   \begin{figure*}
   \centering
   \includegraphics[width=8.9cm]{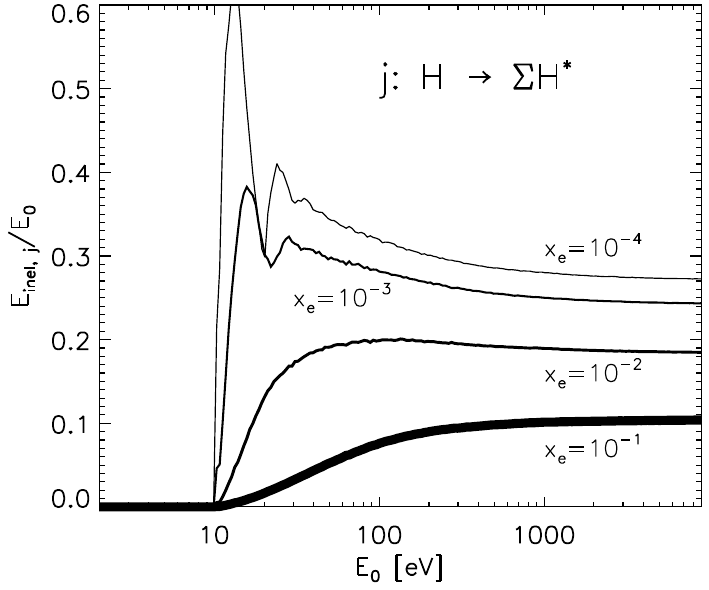}      
   \includegraphics[width=8.9cm]{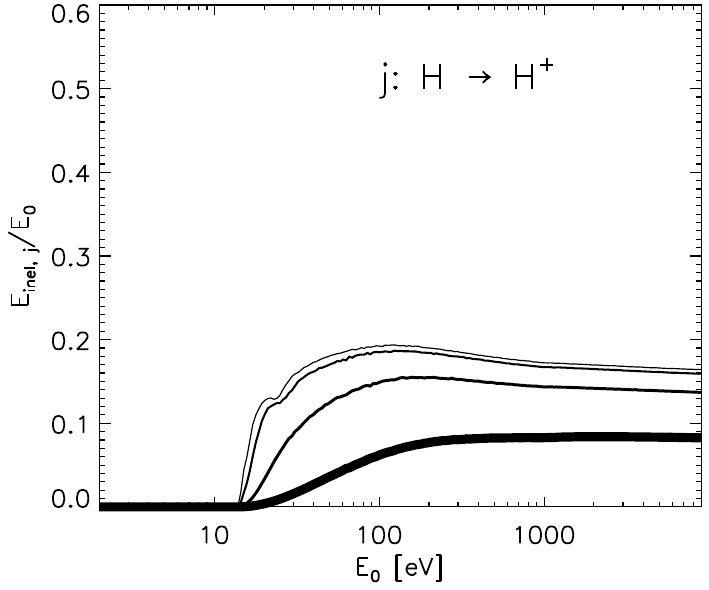} \\
   \includegraphics[width=8.9cm]{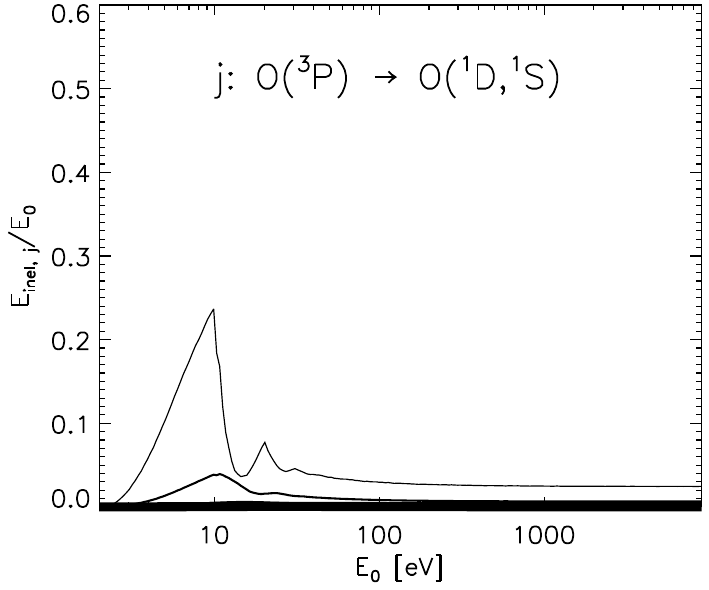}      
   \includegraphics[width=8.9cm]{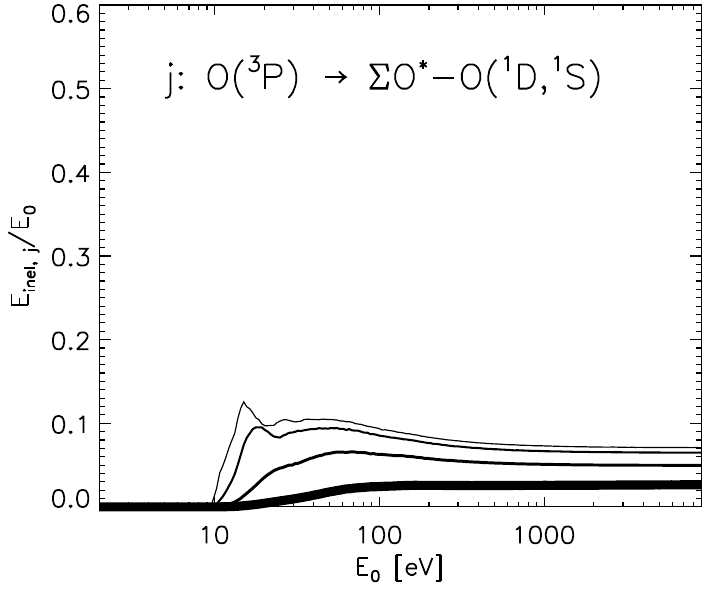} \\
   \includegraphics[width=8.9cm]{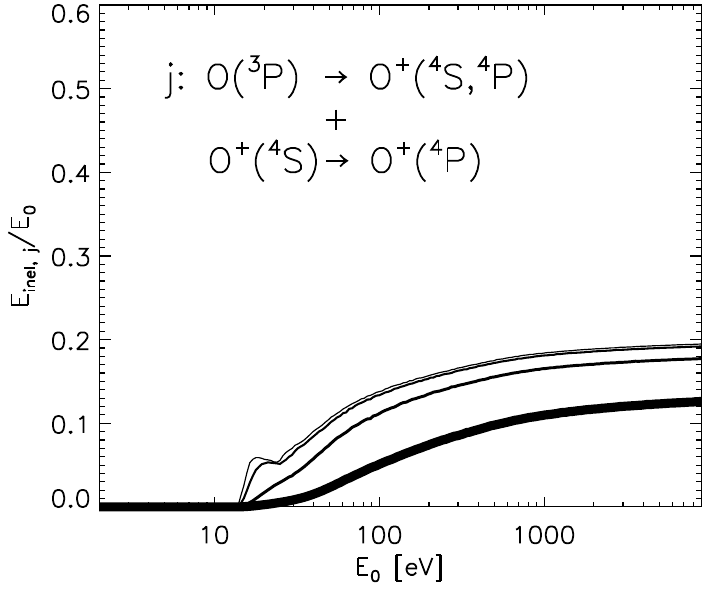}      
   \includegraphics[width=8.9cm]{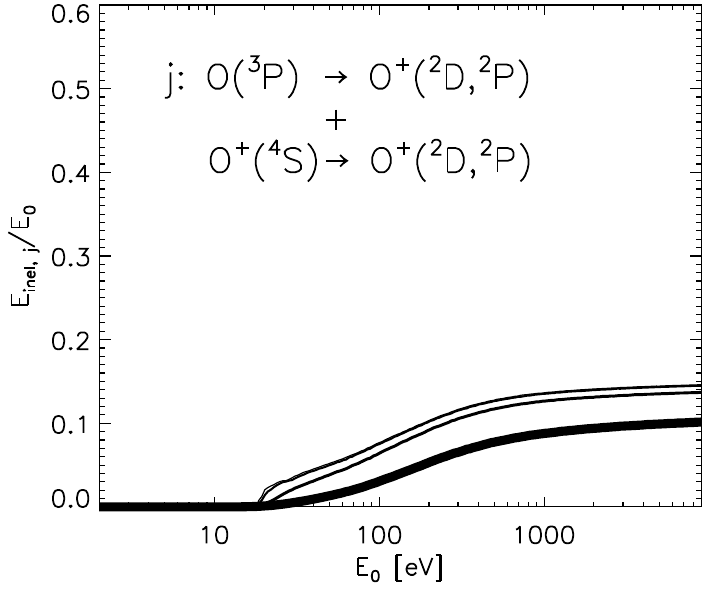} 
   \caption{\label{eta_inelas_steam_fig}
   For an O-H mixture,  $f_{\rm{O_n}/\rm{H_n}}${=}1/2 is the fraction of the photoelectron energy, $E_0$, that goes into various inelastic channels specified by their initial and end levels. Then, H$^*$ and O$^*$ refer to the summation over all excited levels within the corresponding neutral atoms.
   For each panel, 
   the four curves that are displayed (distinguishable by their 
    thickness) refer to fractional ionizations $\log_{10}(x_{\rm{e}})$=$-1$ (thickest), $-2$, $-3,$ and $-4$ (thinnest). 
    Calculations done with the LG90 set of cross sections.
   }
   \end{figure*}

We carried out a systematic investigation of the slowing down of photoelectrons in O-H mixtures.
In the calculations, 
each mixture contains the atoms H, {\hp}, O, and {\op}, plus thermal electrons, $e_{\rm{th}}$.
The number densities of hydrogen and oxygen nuclei are [H$_{\rm{n}}$]=[H]+[\hp] and 
[O$_{\rm{n}}$]=[O]+[\op], respectively, and for the mixture 
[{$X$}$_{\rm{n}}$]=[H$_{\rm{n}}$]+[O$_{\rm{n}}$]. We assume at first that 
H, O, and {\op} are all in their ground levels, but we relax this assumption in the steps detailed in {\S}\ref{o1d_subsec} to explore the lifetime of the metastable {\ooned}. We assume charge neutrality, [\hp]+[\op]=[$e_{\rm{th}}$],   and that [\hp]/[H]=[\op]/[O]. The latter is expected to be a good approximation because charge exchange {\hp}+O$\leftrightarrow$H+{\op} in the atmosphere will tend to rapidly establish the statistical balance 
([\hp]/[H])/([\op]/[O])=9/8. 
This simplification reduces the free parameters in the systematic investigation.
With the above, the relative proportions of  the mixture components are fully specified by the ratio $f_{\rm{O_n}/\rm{H_n}}$=[O$_{\rm{n}}$]/[H$_{\rm{n}}$] and the fractional ionization $x_{\rm{e}}$=[$e_{\rm{th}}$]/[{$X$}$_{\rm{n}}$]. 
We note that  
$f_{\rm{O_n}/\rm{H_n}}$=0 for a pure H gas, =1/2 for the O-H gas resulting from
water dissociation, and {$\infty$} for a pure O gas.
Also, $x_{\rm{e}}$=0 for a neutral gas, and =1, when 
[H]=[O]=0.
After some algebra, the fraction of each atom in the mixture is 
[H]/[{$X$}$_{\rm{n}}$]={($1-x_{\rm{e}}$)}/($1+f_{\rm{O_n}/\rm{H_n}}$), 
[{\hp}]/[{$X$}$_{\rm{n}}$]={$x_{\rm{e}}$}/($1+f_{\rm{O_n}/\rm{H_n}}$),
[O]/[{$X$}$_{\rm{n}}$]=$f_{\rm{O_n}/\rm{H_n}}${($1-x_{\rm{e}}$)}/($1+f_{\rm{O_n}/\rm{H_n}}$), and
[{\op}]/[{$X$}$_{\rm{n}}$]=$f_{\rm{O_n}/\rm{H_n}}${$x_{\rm{e}}$}/($1+f_{\rm{O_n}/\rm{H_n}}$). 
We adopt [{$X$}$_{\rm{n}}$]=10$^{10}$ cm$^{-3}$ in the MC model, although the actual choice has a minor impact on the calculations because, except for the electron-electron 
cross sections, all the other factors in the formulation are independent of the absolute densities. 
Calculations were performed for $f_{\rm{O_n}/\rm{H_n}}$=0, 1/6, 1/2, 3/2, {$\infty$}, and for
$\log_{10}(x_{\rm{e}})$=$-$6, $-$5, $-$4, $-$3, $-$2,  and $-$1, for incident photoelectron energies below 10,000 eV. 

As it slows down, a photoelectron of energy, $E_0$, 
loses a fraction, $E_{\rm{elas}}$, in elastic collisions
and another fraction, $\sum_{j}${$E_{{\rm{inel}},j}$}, in inelastic collisions, where the summation extends over all the available inelastic channels. From energy conservation,  {$E_0$}={$E_{\rm{elas}}$}{$+$}{$\sum_{j}${$E_{{\rm{inel}},j}$}}.\footnote{The equality is exact only if there are no super-elastic collisions in which the photoelectron gains some energy from an excited atom. Because the densities of excited levels are typically very small in exoplanet atmospheres, the equality remains very approximately true.} 
It is usually accepted that  
all of {$E_{\rm{elas}}$} contributes to heating the gas 
and that some of {$\sum_{j}${$E_{{\rm{inel}},j}$}} may also contribute locally. 
Using $E_{\rm{h}}$ to denote the part of $E_0$ that goes into heating, it follows that 
$E_0${$\geq$}{$E_{\rm{h}}$}{$\geq$}{$E_{\rm{elas}}$}. 
The ratio {$E_{\rm{h}}$}/$E_0$ is a heating efficiency, although this  denomination may be confusing as it is used in the literature to describe a variety of related but non-identical concepts.

Figure \ref{eta_yield_steam_fig} summarizes some of the calculations for an O-H mixture with 
$f_{\rm{O_n}/\rm{H_n}}$=1/2, as well as for pure H and pure O. 
The top panel shows $E_{\rm{elas}}$/$E_0$ and the bottom one shows the normalized yield 
$\Phi_{\rm{i}}$/$E_0$. 
The ionization yield, $\Phi_{\rm{i}}$, represents the number of ions produced per injected photoelectron of energy, $E_0$, regardless whether they originated from H, O, or {\op} and of their charge.
Because the MC model considers double ionization for {\othreep}, which leads to {\opp}($^3P$) plus two electrons, $\Phi_{\rm{i}}$ is slightly larger than the number of secondary electrons produced.

We do not discuss the specifics of our calculations for pure H \citep[see e.g.,][for that]{garciamunoz2023} or for pure O.
For the latter, however, we note that they are consistent with what has been reported in past works \citep[see e.g.][]{dalgarnolejeune1971,greenetal1977,slinkeretal1988,kozmafransson1992,victoretal1994}. 
For example, we estimate from Fig. 14 of \citet{dalgarnolejeune1971} and in their calculations, the amount of energy transferred elastically to the thermal electrons for 
$E_0$=50 eV is about 25, 13, 8, 4, and 2.5 eV
for $\log_{10}(x_{\rm{e}})$=$-2$, $-3$, $-4$, $-5,$ and $-6$, respectively. 
 For the same conditions, we obtained the following values: 25.5, 14.9, 8.4, 4.6, and 3.4 eV.

It is seen in Fig. \ref{eta_yield_steam_fig} (top) that $E_{\rm{elas}}$/$E_0$ depends on the gas composition and also on the photoelectron initial energy, 
$E_0,$ and the fractional ionization of the mixture, $x_e$. For $E_0${$<$}10.2 eV, $E_{\rm{elas}}$/$E_0${$\equiv$}1 for pure H, but $<$1 for mixtures that contain O, especially for low $x_{\rm{e}}$. The reason is that the O atom has multiple excitation channels open between 10.2 and 1.96 eV 
(the threshold for {\ooned} excitation) that remain competitive against electron-electron collisions for small-to-moderate $x_{\rm{e}}$.
For {$E_0$}$\gtrsim$100 eV and $x_{\rm{e}}${$\sim$}10$^{-3}$--10$^{-2}$,  $E_{\rm{elas}}$/$E_0$ is typically higher for an O-H mixture with 
$f_{\rm{O_n}/\rm{H_n}}$=1/2 than for pure H. The trend does, however, reverse 
for both lower and higher fractional ionizations.
For pure H and {$E_0$}$\gtrsim$50 eV, the heating efficiency $E_{\rm{elas}}$/$E_0$ becomes nearly independent of the fractional ionization for $x_{\rm{e}}${$\lesssim$}10$^{-4}$. 
In contrast, 
$E_{\rm{elas}}$/$E_0$ drops to $\sim$0.07 
(not shown in Fig. \ref{eta_yield_steam_fig})
in the O-H mixture with $f_{\rm{O_n}/\rm{H_n}}$=1/2 and $x_{\rm{e}}$=10$^{-6}$
because for such small fractional ionizations
excitation into {\ooned} remains efficient at extracting 
energy from the slow photoelectrons.
It is also apparent in Fig. \ref{eta_yield_steam_fig} (bottom) that adding O enhances $\Phi_{\rm{i}}$ and therefore the number of newly created ions
with respect to the case of pure H. 
The reason for this 
is that at energies $\gtrsim$30 eV, the collisions of photoelectrons with O tend to favor its ionization rather than its excitation.

Figure \ref{eta_inelas_steam_fig} is specific to an O-H mixture with $f_{\rm{O_n}/\rm{H_n}}$=1/2, and shows the
ratio $E_{\rm{inel}, j}$/$E_0$ for the various excitation and ionization channels -- except ionization from 
either {\othreep} or
{\opfours} to {\opp}, which remain minor.
We note that excitation is typically more significant than ionization for H, whereas the reverse occurs for O. 
Amongst the O excitation channels, excitation into the metastable {\ooned} plus a smaller contribution to {\oones} stand out as important, especially at low energies. 
Significant amounts of energy go into the simultaneous ionization and excitation of O. 
O$^+$($^4P$) is likely to radiate rapidly to the ground level  {\opfours}.
In contrast, the metastables {\optwod} and {\optwop}
 may return some of their excitation energy to the ground level ion in collisions with the other gases in the atmosphere.
Interestingly, 
most of the newly created ions are produced through O ionization rather than through H ionization, even though the ratio of number densities in these calculations is [O]/[H]=1/2.

\begin{figure*}[h]
   \centering
   \includegraphics[width=8.5cm]{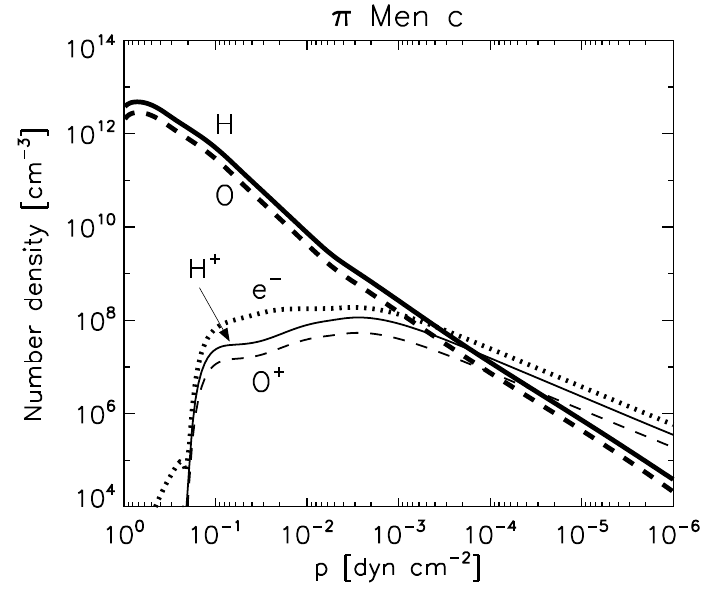} 
   \includegraphics[width=8.5cm]{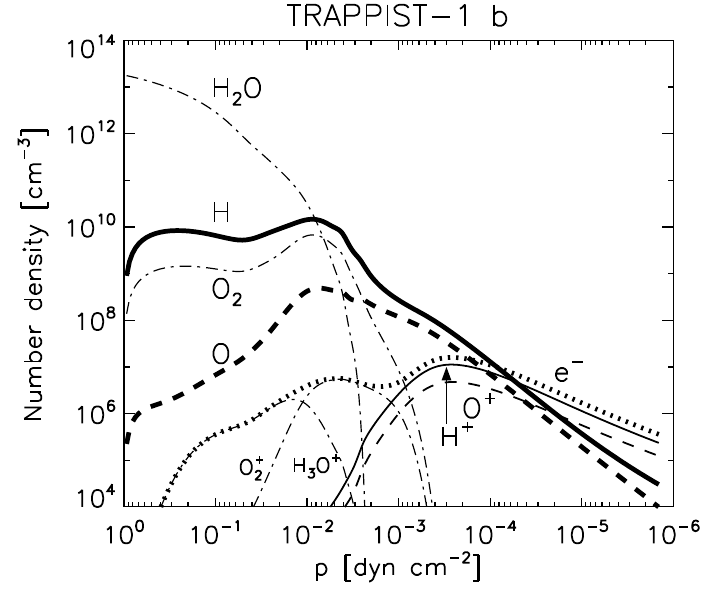} \\
   \includegraphics[width=8.5cm]{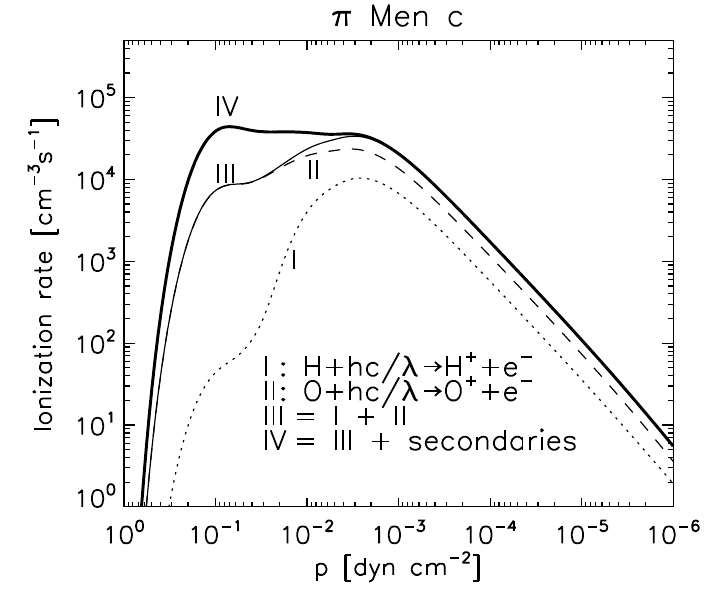} 
   \includegraphics[width=8.5cm]{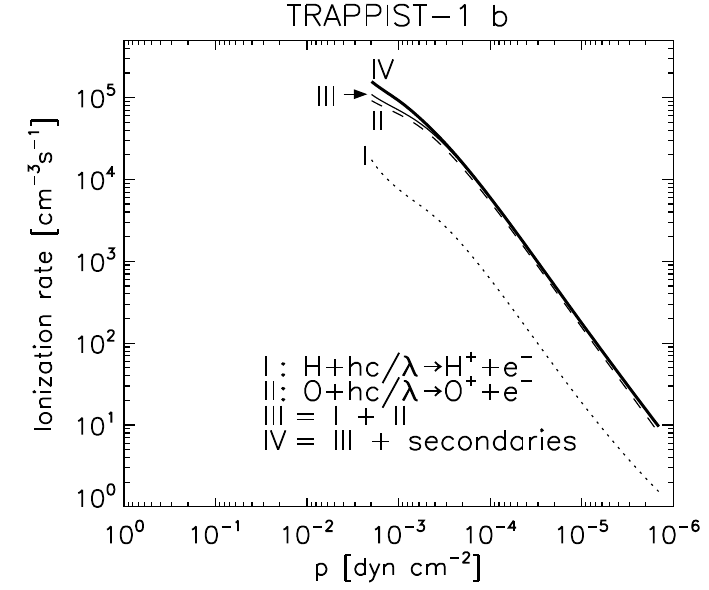} \\
   \includegraphics[width=8.5cm]{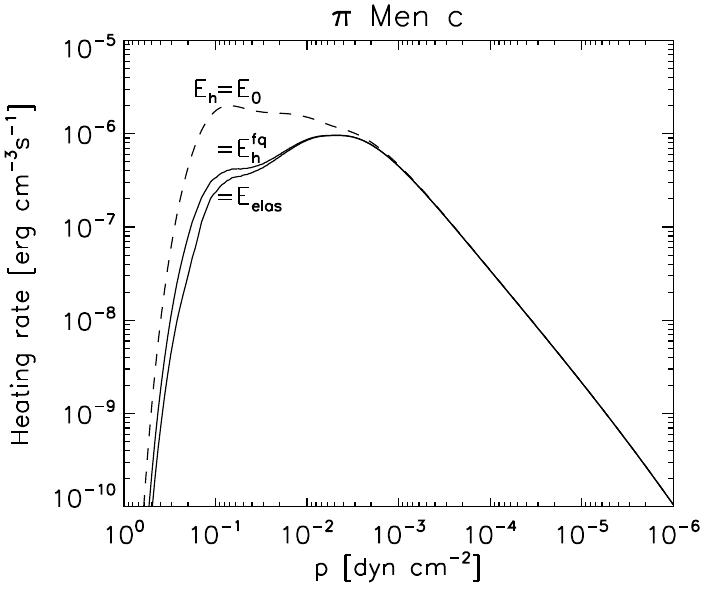} 
   \includegraphics[width=8.5cm]{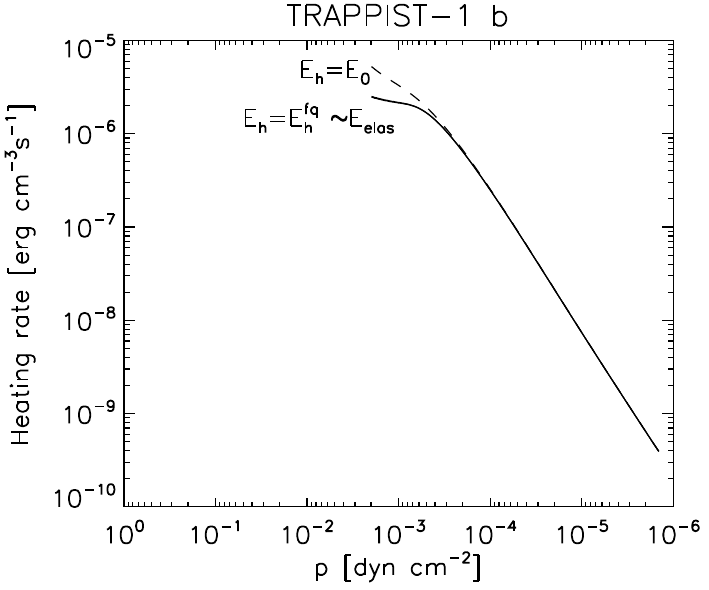} 
   \caption{\label{profilesnd_fig}
   Some properties of the atmospheres of {\pimenc} and {\trappistoneb}. 
   Top: 
   Adopted number densities of H, {\hp}, O, {\op} and $e_{\rm{th}}$. See text for details. 
   For {\trappistoneb}, the densities of additional molecules ({\htwoo}, {\otwo}, {\hthreeop}, {\otwop}) that remain relatively abundant up to 10$^{-3}$ dyn cm$^{-2}$ are also shown. 
   Middle: Production rates of primary and secondary ions. 
   Bottom: Heating rates in the no- and full-quenching limits (see text), and under the assumption that photoelectrons transfer all their kinetic energy into heat.
   }
   \end{figure*}

\section{\label{application_sec} Application to {\pimenc} and {\trappistoneb}}

The heating and ionization of an atmosphere are partly dictated by the 
interaction of the photoelectrons with the gas
while they slow down to thermal energies.
That information is encoded in $E_{\rm{elas}}$/$E_0$ and $\Phi_{\rm{i}}$. 
A more complete description 
requires, in addition, tracking the fate of the excited and ionized levels produced by the photoelectrons. 
For example, the energy $E_{\rm{h}}$ that goes into heat will exceed $E_{\rm{elas}}$ if the
population of excited levels produced by the photoelectrons is rapidly quenched (collisionally deexcited) to a lower-energy level.
The problem is complex and non-local, and involves solving 
self-consistently 
the equations of mass, momentum and energy conservation, and radiative transfer in the atmosphere.

Here, we focus on the simpler problem 
of estimating the heating and ionization that occurs locally. 
These quantities depend on 
the local density, composition and fractional ionization of the gas in ways that are described below. 
We devote the bulk of our study to the exoplanets  {\pimenc} \citep{gandolfietal2018,huangetal2018} and 
{\trappistoneb} \citep{gillonetal2017},
 whose atmospheres might prove to be water-rich,
 according to some analyses
 \citep{garciamunozetal2021,agoletal2021}. 

For {\pimenc}
($M_{\rm{p}}$/$M_{\oplus}$=4.52{$\pm$}0.81, $R_{\rm{p}}$/$R_{\oplus}$=2.06{$\pm$}0.03; orbital distance $a$=0.067 AU; \citet{gandolfietal2018}), we take the number densities
in the upper atmosphere from a model published by \citet{garciamunozetal2021} that assumes $f_{\rm{O_n}/\rm{H_n}}${$\approx$}1/2
(their case 6). 
{\trappistoneb} 
($M_{\rm{p}}$/$M_{\oplus}$=1.374{$\pm$}0.069, $R_{\rm{p}}$/$R_{\oplus}$=1.116$^{+0.014}_{-0.012}$; $a$=(1.154{$\pm$}0.010)$\times$10$^{-2}$ AU; \citet{agoletal2021}) is part of a system of seven known planets, some of which orbit in the habitable zone of their ultracool dwarf host star.
For {\trappistoneb, we used} the same photochemical-hydrodynamical model as in \citet{garciamunozetal2021} to predict the number densities. 
We prescribe at the bottom boundary of the model (pressure of 1 dyn cm$^{-2}$) that the atmosphere is pure {\htwoo} and let the model evolve its composition at higher altitudes. We adopted the TRAPPIST-1 spectrum published by \citet{wilsonetal2021}, 
which shows that the far-ultraviolet emission of the star (a wavelength range usually key in the photodissociation of molecules) is rather faint compared to the XUV emission.

Figure \ref{profilesnd_fig} (top) shows the number densities of H, {\hp}, O, {\op} and $e_{\rm{th}}$ for both planets, which we adopt 
throughout our analysis as shown there. 
At pressures of {$\gtrsim$}10$^{-3}$ dyn cm$^{-2}$, the chemistry of {\trappistoneb} becomes particularly interesting because the weak photodissociation at far-ultraviolet wavelengths facilitates the survival of molecules.
Indeed, {\htwoo} and {\otwo} are abundant in that region
amongst the neutrals, with
{\hthreeop} and {\otwop} amongst the ions. 
At lower pressures, the {\htwoo} molecule is no longer shielded by the gas column overhead and its abundance drops rapidly, and so does the abundance of the other molecules.
For {\trappistoneb}, we focus our analysis on pressures {$\lesssim$}10$^{-3}$ dyn cm$^{-2}$, for which the chemistry remains mainly atomic.
We recall that the model in 
\citet{garciamunozetal2021} does not take into account the non-thermal 
effects described here or, equivalently, that it assumes
$E_{\rm{h}}$/$E_0${$\equiv$}1 and $\Phi_{\rm{i}}${$\equiv$}0. 
Our analysis below is therefore not self-consistent 
because it does not allow for feedbacks that might modify the atmosphere. 
Even with these caveats, which will be fixed in follow-up work, 
the analysis clearly demonstrates and partially quantifies the importance of non-thermal effects in water-rich atmospheres.

We calculated the production rates for primary photoelectrons from 
${\rm{H}}+hc/\lambda\rightarrow{{\rm{H}^+}}+e^{-}$ and ${\rm{O}}+hc/\lambda\rightarrow{{\rm{O}^+}}+e^{-}$. We note that $hc$ is the product of Planck's constant and the speed of light, and $\lambda$ is the wavelength of the ionizing photon.
They are shown in Fig. \ref{profilesnd_fig} (middle) as curves I and II, while curves III show the total production rates of primary photoelectrons (=I+II). The
O atom contributes most of the primary photoelectrons, even though its abundance is about half that of H. 
The reason for this is that the cross section for O photoionization 
is generally larger than the cross section for H photoionization. 
The disparity in the production rates between the O and H atoms is exacerbated in the case of 
{\trappistoneb} by the large fraction of XUV photons emitted by its host star at the shortest wavelengths.

\begin{figure*}[t]
   \centering
   \includegraphics[width=8.9cm]{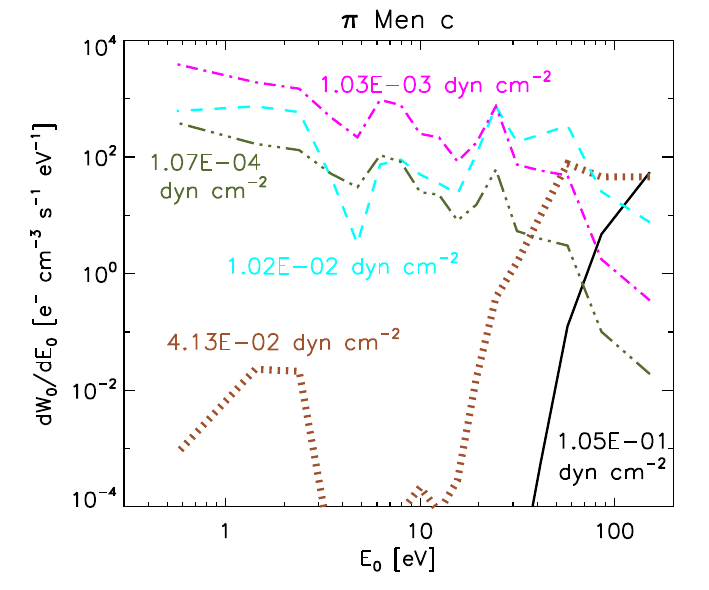}  
   \caption{\label{dWE0_pimenc_fig}
   Energy spectra of the nascent photoelectrons produced through photoionization of O and H atoms at different levels in the atmosphere of {\pimenc} for    
   $f_{\rm{O_n}/\rm{H_n}}$=1/2. The quoted numbers in the plot represent the pressure levels. 
   }
   \end{figure*}

Each primary photoelectron of energy $E_0$ ($\approx${$hc$/$\lambda$}{$-$}13.6 eV for both H and O, where for consistency with the photochemical-hydrodynamical model it is assumed that O photoionization results in ground-level {\op})
produces a number of secondary electrons that depends on $E_0$ and 
the local composition and fractional ionization of the atmosphere. 
We used the number densities of H, {\hp}, O, {\op}, and $e_{\rm{th}}$ from Fig. \ref{profilesnd_fig}, plus the energy spectrum of the primary photoelectrons, to calculate the yield  $\Phi_{\rm{i}}$ and, ultimately, the net ionization rate at each altitude  following conventional methods \citep{garciamunoz2023}. 
Figure \ref{dWE0_pimenc_fig} shows the energy spectrum of the primary photoelectrons
in the atmosphere of {\pimenc} over a range of pressures. 
At low pressures, photoionization of the gas by photons of wavelengths near the 
Lyman continuum threshold results mainly in low-energy photoelectrons.
Deeper in the atmosphere, where only very energetic photons penetrate, 
the nascent photoelectrons become increasingly energized.

\begin{figure*}[b]
   \centering
   \includegraphics[width=8.5cm]{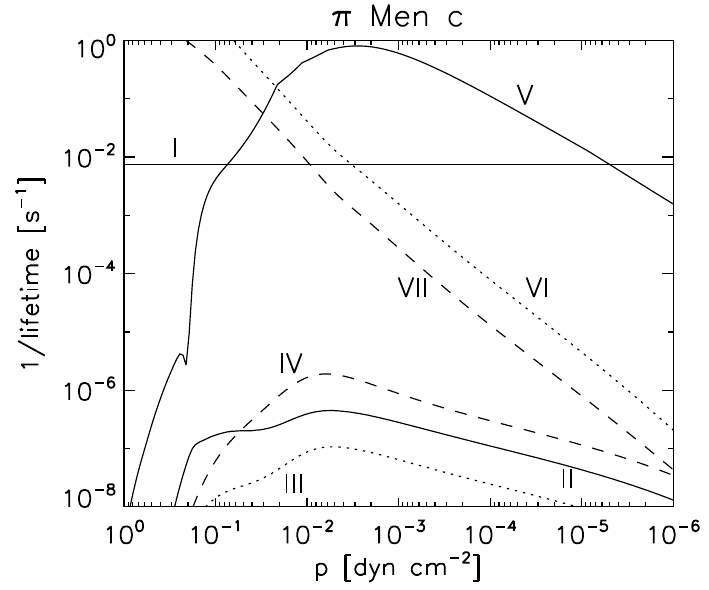} 
   \includegraphics[width=8.5cm]{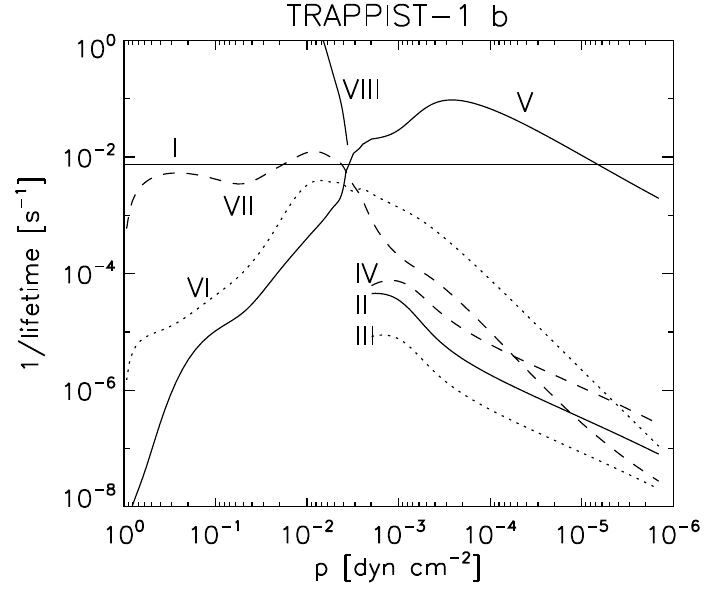} \\
   \caption{\label{lifetimes_fig}
   Some lifetimes for {\ooned}. I: Radiative emission, {\ooned}$\rightarrow${\othreep}+$h\nu$; 
   II, III, and IV: 
   Quenching $e'$+{\ooned}$\rightarrow${$e''$}+{\othreep},  
   excitation $e'$+{\ooned}$\rightarrow${$e''$}+{\oones}, and
   ionization $e'$+{\ooned}$\rightarrow${$e''$}+{$e'''$}+{\op} by photoelectrons, 
   respectively;
   V, VI and VII: 
   quenching by thermal electrons {$e_{\rm{th}}$}+{\ooned}$\rightarrow${$e_{\rm{th}}$}+{\othreep}, 
   quenching by oxygen atoms {\othreep}+{\ooned}$\rightarrow${\othreep}+{\othreep}, 
   quenching by hydrogen atoms {H(1)}+{\ooned}$\rightarrow${H(1)}+{\othreep}, respectively;
   VIII (for TRAPPIST-1 b only): chemical transformation {\ooned}+{\htwoo}$\rightarrow${2OH}.   
   }
   \end{figure*}

\subsection{Ionization and heating rates}

The total production rates, that is, primary plus secondary electrons, are shown in Fig. \ref{profilesnd_fig} (middle) as curves IV. 
The main finding for {\pimenc} is that the photoelectrons enhance the ionization rate by a factor of a few at 
{$p$}$\gtrsim$10$^{-2}$ dyn cm$^{-2}$. This region of the atmosphere is key because it defines where the stellar energy is deposited and the gas becomes accelerated to escape the planet. 
A change in the net ionization rate will surely have  implications on these important aspects.
For {\trappistoneb}, at {$p$}$\sim$2$\times$10$^{-3}$ dyn cm$^{-2}$ the contribution of photoelectrons to the overall ionization rate is about 40{\%}. It is justifiable to assume that as for {\pimenc}, their relative contribution will 
increase towards the deeper atmosphere. 
The high abundance of molecules there (in particular {\htwoo}) may, however, introduce specific features in the ionization balance that are worth investigating in the future.

For the conditions encountered in the hydrogen-dominated 
atmospheres of close-in exoplanets at pressures of $\lesssim$1 dyn cm$^{-2}$, the excited levels of hydrogen will likely radiate away their energy, therefore not contributing to the local heating of the atmosphere \citep[see e.g., ][for the specific conditions of the ultrahot Jupiter HAT-P-32 b]{garciamunoz2023}.\footnote{There are exceptions to this, though. For another ultrahot Jupiter, KELT-9 b, the excited level H(2) becomes rapidly photoionized by near-ultraviolet radiation and thus contributes to heating.}
This occurs even if the lines in the Lyman series, which connect with the ground level H(1$s$), are optically thick. Also, the excited level H(2$s$), which does not connect with H(1$s$) through an optically allowed transition, will typically radiate by conversion first into 
H(2$p$) via collisions with protons.
Similar arguments should be valid for most of the excited levels of O and {\op}, except (possibly)  for
the metastables O($^1D$), O($^1S$), {\op}($^2D$), and {\op}($^2P$), which have long radiative lifetimes and may become quenched before they radiate.

When considering the contribution of the  metastables to heating, two limits can be identified.
In the no-quenching limit, which should be valid when collisions are rare and quenching is inefficient,
only the kinetic energy transferred by the photoelectrons to the thermal electrons and heavy particles in elastic collisions contributes to heating, 
$$
E^{\rm{nq}}_{\rm{h}}=E_{\rm{elas}}.
$$
Alternatively, in the full-quenching limit, which should apply when many deexcitation collisions occur before the atom radiates, 
$$
E^{\rm{fq}}_{\rm{h}}=E_{\rm{elas}}
+E_{\rm{inel},O(^3P)\rightarrow O(^1D)}
+E_{\rm{inel},O(^3P)\rightarrow O(^1S)} + 
$$
$$
+E_{\rm{inel},O(^3P)\rightarrow O^+(^2D)} \frac{(16.93-13.61)}{16.93}
$$
$$
+E_{\rm{inel},O(^3P)\rightarrow O^+(^2P)} \frac{(18.63-13.61)}{18.63} 
$$
$$
+E_{\rm{inel},O^+(^4S)\rightarrow O^+(^2D)} +E_{\rm{inel},O^+(^4S)\rightarrow O^+(^2P)}.
$$
The latter assumes that when simultaneous ionization-excitation occurs, 
quenching converts into heat only the excitation energy above the ground level of the ion. 

We calculated the heating rate in the atmosphere as follows:
$$
\mathcal{Q} = \int ( [{\rm{H}}] \sigma^{\rm{pi}}_{\rm{H}}(\lambda) +
[{\rm{O}}] \sigma^{\rm{pi}}_{\rm{O}}(\lambda)) \mathcal{F_{\star}}(\lambda)  \left( 1-{\lambda}/{\lambda_{\rm{Ly,c}}} \right)
E_{\rm{h}}/E_0 d{\lambda},
$$
where $\sigma^{\rm{pi}}$ stands for the photoionization cross section, $\lambda_{\rm{Ly,c}}$ is the
Lyman continuum threshold (to practical effects, the H and O atoms have the same ionization potential), 
and the stellar radiation flux $\mathcal{F_{\star}}(\lambda)$ is expressed in energy units. 
The results for $E_{\rm{h}}$=$E^{\rm{nq}}_{\rm{h}}$ and $E_{\rm{h}}$=$E^{\rm{fq}}_{\rm{h}}$
are shown in Fig. \ref{profilesnd_fig} (bottom) as two different curves. They should bracket the actual heating rate in the atmosphere because in general the metastables lose a part of their energy to radiation and another part to collisional deexcitation.
For reference, the heating rate calculated for $E_{\rm{h}}$/$E_0${$\equiv$}1, as in the photochemical-hydrodynamical model of \citet{garciamunozetal2021}, is also shown.

One conclusion from this figure is that metastable quenching may enhance $E_{\rm{h}}$ over $E_{\rm{elas}}$ by minor amounts in the cases under study here. 
More importantly perhaps, 
the figure shows that neglecting photoelectrons may overestimate the heating rate by a factor of a few. 
A more accurate determination of this requires considering possible feedbacks in the atmosphere.

\subsection{\label{o1d_subsec} Metastables {\ooned} and {\oones}}

The two lowermost excited levels of atomic oxygen, {\ooned} and {\oones}, lie at only 1.96 and 4.18 eV above the ground-level {\othreep}. They are metastable, with total radiative lifetimes of 134 and 0.75 s, respectively. They form through various processes, including collisional excitation by thermal and non-thermal electrons, photodissociation of {\htwoo} or {\otwo}, and dissociative recombination of {\otwop}. As both {\ooned} and {\oones} may act as efficient coolants by emitting in the optical and ultraviolet, it is relevant to ask whether collisions with photoelectrons might shorten their effective lifetimes, thereby potentially modifying the energy budget of the atmosphere. 

To answer that question, we have gathered the basic set of cross sections shown in Fig. \ref{xs_others_fig} for 
collisions of {\ooned}. They include
deexcitation into {\othreep}, 
excitation into {\oones}  and ionization into {\opfours}. Following the small abundance approximation
described in \citet{garciamunoz2023}, we determined the lifetimes for {\ooned} in deexcitation, excitation, and ionization collisions with photoelectrons. We know that
{\ooned} photoionizes at wavelengths $\lesssim$830 {\AA},  
a spectral region effectively shielded by the dominating neutral gases. 
Our calculations confirm that 
{\ooned} photoionization is slow compared to spontaneous emission or quenching by thermal electrons or heavy particles and it is thus omitted from the discussion.

The results of the calculations are shown in Fig. \ref{lifetimes_fig} for {\pimenc} and {\trappistoneb}.
For comparison, we also show the {\ooned} radiative lifetime, 
the quenching lifetimes in collisions with thermal electrons 
(rate coefficients based on \citet{barklem2007}), 
O atoms \citep{jamiesonetal1992},
and H atoms \citep{kremsetal2006}.
For {\trappistoneb}, we also give the lifetime for chemical reaction with {\htwoo} molecules 
(rate coefficient from \citet{burkholderetal2019}).
Clearly, the lifetime of {\ooned} is negligibly affected by photoelectron collisions. 
Rather, deexcitation collisions with O and H atoms and chemical reactions with {\htwoo} control the {\ooned} lifetime in the mostly neutral atmosphere; while deexcitation collisions with thermal electrons take the control when the fractional ionization reaches a few times 10$^{-3}$.
The situation is reminiscent of the modeling of stellar atmospheres, where it is thought that the population of oxygen levels is affected by inelastic collisions with both hydrogen atoms and thermal electrons, with the importance of each type of collision depending on the stellar effective temperature \citep{allendeprietoetal2003,barklem2018}.

The cross sections for the deexcitation, excitation, and ionization of {\oones} in collisions with photoelectrons are comparable or smaller than those for {\ooned} \citep{barklem2007,tayalzatsarinny2016}. 
Given that the radiative lifetime of {\oones} is two orders of magnitude shorter than that of {\ooned}, it turns out that 
the lifetime of {\oones} is also negligibly affected by photoelectron collisions. The corresponding calculations of lifetimes are not presented.

\section{\label{summary_sec} Summary and perspectives}

We simulated the slowing down of photoelectrons in O-H mixtures, determining the yields for each elastic and inelastic collisional channel and the heating efficiencies. 
Our work extends the calculations done for hydrogen-dominated atmospheres by \citet{cecchi-pestellinietal2006,cecchi-pestellinietal2009} 
and \citet{shematovichetal2014} to atmospheric compositions that may be more relevant in the characterization of small exoplanets.

We find that the fraction $E_{\rm{elas}}$/$E_0$ of the photoelectron energy that goes into elastic collisions (and thus into heat) for an O-H mixture is not very different from what is predicted for pure H. The exception to this behaviour occurs at fractional ionizations $x_{\rm{e}}${$<<$}10$^{-4}$, as in these conditions 
$E_{\rm{elas}}$/$E_0$ can be significantly lower than for pure H.
The difference arises from the fact that excitation in the O atom remains possible down to energies of about 2 eV, unlike the case of the H atom. 
We also find that for a given rate of primary photoelectrons, the total (=primary + secondary) number of electrons produced in an O-H mixture is  higher than for pure H.
 
 We have investigated the significance of non-thermal effects in the upper atmospheres of two possibly water-rich exoplanets.
 We predict that the atmospheres that contain abundant O will ionize faster than those that are composed of pure H. The
 reason for this is that the O atom has cross sections for photoionization and photoelectron-collision ionization that are notably larger than those for H. Faster ionization of O, along with subsequent ionization of H through charge exchange in the atmosphere, will surely have an impact on the detectability of H atoms escaping the atmospheres of water-rich exoplanets through {\lalpha} transmission spectroscopy. 
 This finding reinforces the idea postulated in \citet{garciamunozetal2020} that the published detections of 
 {\lalpha} absorption are correlated with bulk densities of the planets below 2--3 g cm$^{-3}$, which is likely indicative of low-metallicity atmospheres. At the same time, the non-detections are consistent with higher bulk densities consistent with high-metallicity atmospheres.
 Subsequent detections of {\lalpha} absorption for
  HAT-P-11 b ($\rho_{\rm{p}}${$\sim$1.6 g cm$^{-3}$}; \citet{ben-jaffeletal2022}) and
 HD 63433 c ({$\sim$2.1 g cm$^{-3}$}; \citet{zhangetal2022}),
and non-detections for HD 63433 b 
({$\sim$3 g cm$^{-3}$}; \citet{zhangetal2022}), 
K2-25 b ({$\sim$3.3 g cm$^{-3}$}; \citet{rockcliffeetal2021}), 
GJ 9827 b ({$\sim$7.5 g cm$^{-3}$}; \citet{carleoetal2021}) 
are consistent with the idea, 
even though the detection of {\lalpha} absorption may depend on other factors as well such as the overall stellar XUV flux or the stellar wind strength \citep[e.g.,][]{shaikhislamovetal2020}.

 We argue that omitting the effect of photoelectrons can artificially boost the heating of water-rich atmospheres by a factor of a few in the layers where most of the high-energy stellar photons are deposited and the atmospheric outflow is established. 
 The quenching of metastable levels of O and {\op} produced by the photoelectrons does not appreciably change these findings. 
 A weaker heating rate will surely extend the long-term stability of
 water-rich atmospheres.
 
As we strive to understand the nature of water-rich exoplanets, the models that we are developing ought to explore the main atmospheric processes that play a role in them. 
That mission has not been accomplished yet and in what follows, we list some of the caveats of the present work. Our treatment of the collisional-radiative processes remains incomplete as it does not consider the excitation of O and {\op} in collisions with thermal electrons and other heavy particles, nor diffuse continuum radiation that might be important at the altitudes where the atmosphere transitions from optically thick to optically thin at XUV wavelengths. 
Additionally, 
a change in the heating efficiency of the atmosphere attributable to non-thermal processes will modify the thermal cooling of the gas and, inversely, changes in the temperature due thermal cooling will affect the overall structure of the atmosphere and its fractional ionization -- thereby modifying the effect of non-thermal processes. 
Our exploration of the {\trappistoneb} system reveals that the conversion of {\htwoo} into its atomic components depends strongly on the specifics of the stellar spectrum. We have not explored how efficiently the photoelectrons produced by deposition of XUV photons will dissociate and ionize the {\htwoo} molecule in the transition region between the lower and upper atmospheres, as this calls for additional work. 
Lastly, we have not explored the feedbacks between thermal and non-thermal processes and the hydrodynamics of the flow. The combined effects of these processes may contribute constructively or destructively in ways that need to be quantified, and we will explore these issues in due course.


\clearpage


\clearpage


\clearpage

\begin{appendix}

\section{\label{xs_appendix} Collisional and radiative properties of the model}

The MC model for the slowing down of photoelectrons in a pure H gas is described in \citet{garciamunoz2023}. 
It considers: 
one elastic channel for electron-electron collisions; one elastic channel for electron-H collisions; six inelastic channels for electron collisions with the H atom in its ground level that lead to excitation into levels with principal quantum numbers up to five and into a pseudo-level representative of higher-energy levels; one inelastic channel for ionization. The model can optionally treat the excitation, deexcitation and ionization of various excited levels of the H atom.

In what follows, we describe the expansion of the MC model to take into account collisions of photoelectrons with O and {\op}. 
Typically, a calculation involves one elastic channel for electron-{\othreep} collisions, 66 inelastic channels for collisions with {\othreep} and four inelastic channels for collisions with {\opfours}.

\subsection{Collisions with the ground-level atoms {\othreep} and 
{\opfours}. Sensitivity to different sets of cross sections}

Momentum-transfer cross sections for elastic collisions
of the ground-level atom {\othreep} have been reported in the literature, for instance: \citet{itikawa1978} and \citet{tayalzatsarinny2016}. Both sets are consistent and we adopted the most recent one, which covers a broader range of energies. 
\citet{lahergilmore1990} compiled a set
(hereafter the LG90 set) of inelastic cross sections for {\othreep}
that remains comprehensive to date. 
After correction for auto-ionization, it includes 
61 channels for excitation within the neutral atom, 
one channel for ionization into the ground level ion {\opfours}, 3 channels for ionization-excitation into {O$^+$}($^2D^{\rm{o}}$, $^2P^{\rm{o}}$, $^4P$), 
and one channel for double ionization into {\opp}($^3P$).  
The latter is included for completeness, but given its small cross sections, 
the effect of double ionization is minor in our calculations.  

There have been more recent calculations of some of the inelastic cross sections in the LG90 set
\citep[e.g.,][]{barklem2007, tayalzatsarinny2016}, 
which motivated us to assess the sensitivity to them of our results. 
Our LG90-TZ16 set of cross sections relies on the LG90 set but, when possible, we replaced the older determinations 
by the \citet{tayalzatsarinny2016} calculations. 
In particular, we adopted, from the latter work, the total cross sections for single ionization, which include auto-ionization.  
For the relative branching ratios into $^4S^{\rm{o}}$, $^2D^{\rm{o}}$, $^2P^o$, and $^4P$,  
we implemented the branching ratios from the LG90 set. 
For the LG90-TZ16 set, we also updated the cross sections for the channels with end levels (excitation energy in parenthesis, in eV): 
$2p^4$ $^1D$ (1.96), 
$2p^4$ $^1S$ (4.18), 
$2p^3(^4S^{\rm{o}})3s$ $^5S^{\rm{o}}$ (9.14), 
$2p^3(^4S^{\rm{o}})3s$ $^3S^{\rm{o}}$ (9.51),
$2p^3(^4S^{\rm{o}})3p$ $^5P$  (10.73),
$2p^3(^4S^{\rm{o}})3p$ $^3P$  (10.98), 
$2p^3(^4S^{\rm{o}})3d$ $^3D^{\rm{o}}$ (12.08),
$2p^3(^2D^{\rm{o}})3s$ $^3D^{\rm{o}}$ (12.53),
$2p^3(^2P^{\rm{o}})3s$ $^3P^{\rm{o}}$ (14.11), and 
$2s 2p^5$  $^3P^{\rm{o}}$      (15.65).
The update includes excitation 
into {\ooned}, which has the largest cross section of all excitation channels.
In particular, 
we assumed for the LG90-TZ16 set that excitation into 2$p^3$($^2P^o$)3$s$ $^3P^o$ 
occurs without ionization; we also assumed that excitation into 2$s$2$p^5$ $^3P^{\rm{o}}$ occurs with the excited level always auto-ionizing and therefore its excitation cross section is zero.
We extracted the data from \citet{tayalzatsarinny2016}
by scanning the corresponding graphs. 
As needed,  
the 
\citet{lahergilmore1990} data were used 
at higher energies.

We additionally considered inelastic collisions
of the photoelectrons with the ground level ion O$^+$($^4S^{\rm{o}}$), including ionization into O$^{2+}$($^3P$)  \citep[cross sections from][]{belletal1983} and excitation into 
{\op}($^2D^{\rm{o}}$, $^2P^{\rm{o}}$, $^4P$) \citep[cross sections from][]{itikawaetal1985}. The 
excitation cross sections were extrapolated beyond their stated range of validity, although this should have a negligible effect as ionization seems to dominate at high energies. 
The cross sections for collisions with O$^+$($^4S^{\rm{o}}$) are presented in Fig. \ref{xs_others_fig} (top).

For the probability distribution function of the energy of  post-collision electrons, 
we used the formulation described in \citet{garciamunoz2023}. 
In particular, we implemented $\bar{E}$({\othreep})=13.3 eV 
as an average of the equivalent factors reported in 
 \citet{itikawaichimura1990} for the single ionization channels and,
for simplicity, we adopted the same value for the double ionization channel.
In double ionization collisions, the MC model determines the energy of one of the post-collision photoelectrons directly from the probability distribution function $P$($E''$; $E'$); it is then assumed that the other two post-collision photoelectrons take each of them half of the remaining energy. 
When needed, we took 
\citep{kozmafransson1992}, $\bar{E}$({\ooned})=0.6$\times$IP({\ooned})=7 eV and 
$\bar{E}$({O$^+$($^4S^{\rm{o}}$)})=0.6$\times$IP({O$^+$($^4S^{\rm{o}}$)})=21 eV.

We carried out the calculations presented in 
Figs. \ref{eta_yield_steam_fig} and \ref{eta_inelas_steam_fig}
with both the LG90 and LG90-TZ16 sets of cross sections. We found that the calculated 
$E_{\rm{elas}}$/$E_0$ and $\Phi_{\rm{i}}$ differed by typically less than 10{\%}. This amount that serves as a reasonable guess for the accuracy of our calculations, which should be dominated by the accuracy of the cross sections -- and not by the solution method. 
The discrepancies are largely dictated by the dominant inelastic channels, namely: 
excitation into O($^1D$) below $\sim$20 eV and ionization at higher energies. 
The rest of calculations presented in this work were done with the LG90 set.

\subsection{Collisions with {\ooned}}

\begin{figure*}[h]
   \centering
   \includegraphics[width=15.cm]{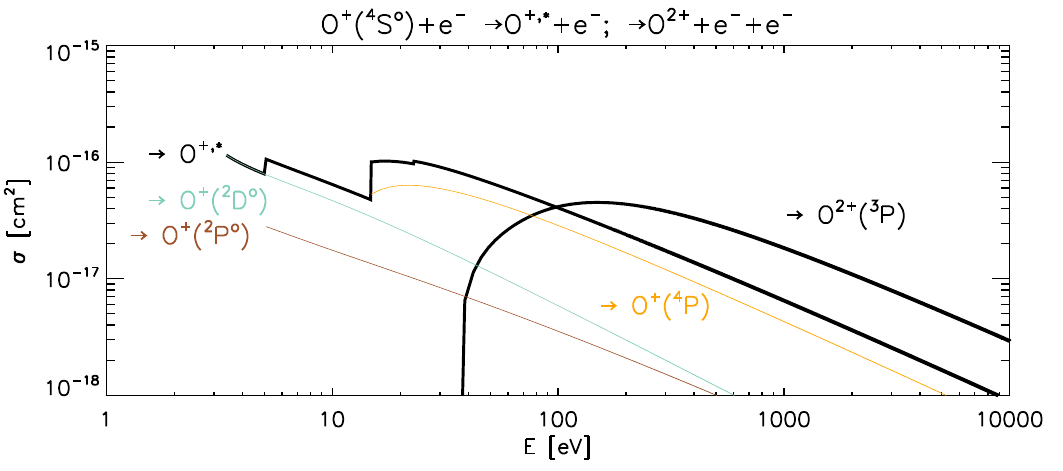}  \\
   \includegraphics[width=15.cm]{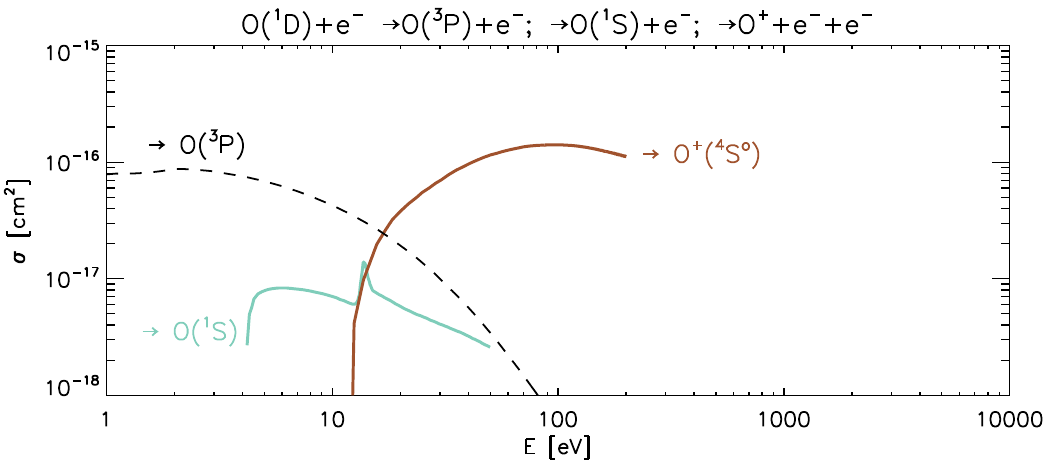}  \\
   \caption{\label{xs_others_fig}
   Adopted cross sections for the excitation and ionization of O$^+$($^4S^{\rm{o}}$) (top) and for (de)excitation and ionization of {\ooned} (bottom). Note: in the top panel, the black curve labeled O$^{+,*}$ includes all the excitation channels that are shown in the graph.
   }
   \end{figure*}

We considered the superelastic channel for deexcitation into {\othreep}, the cross sections which were obtained from the application of detailed balancing to the excitation cross sections, along with two inelastic channels for excitation into {\oones} and single ionization into O$^+$($^4S^{\rm{o}}$). We took the inelastic cross sections from \citet{tayalzatsarinny2016} 
and confirmed them with Fig. 1 of \citet{barklem2007}, finding that it is probably safe to neglect other excitation channels. The momentum-transfer cross sections for 
{\ooned} are comparable in magnitude to those for {\othreep}, and given the much lower abundances
expected for {\ooned,} the corresponding collisional channel needs not be included.
The cross sections for collisions with {\ooned} are presented in Fig. \ref{xs_others_fig} (bottom).

\end{appendix}


\begin{thebibliography}{}

\bibitem[Agol et al.(2021)]{agoletal2021}
{Agol}, E., {Dorn}, C., {Grimm}, S.~L., {Turbet}, M., {Ducrot}, E., et al.\ 2021,
Planetary Science Journal, 2:1,
DOI: 10.3847/PSJ/abd022.

\bibitem[Allende Prieto et al.(2003)]{allendeprietoetal2003}
{Allende Prieto}, C., 
{Lambert}, D.~L., 
{Hubeny}, I. \& {Lanz}, T.\ 2003,
\apjs, 147:363-368,
DOI: 10.1086/375213.

\bibitem[Ashihara(1978)]{ashihara1978}
{Ashihara}, O.\ 1978,
\icarus, 35:369-384,
DOI: 10.1016/0019-1035(78)90089-1

\bibitem[Barklem(2007)]{barklem2007}
Barklem, P.S.\ 2007,
\aap, 462:781-788, 
DOI: 10.1051/0004-6361:20066341.

\bibitem[Barklem(2018)]{barklem2018}
{Barklem}, P.~S.\ 2018,
\aap, 610:A57,
DOI: 10.1051/0004-6361/201731968.

\bibitem[Bell et al.(1983)]{belletal1983}
Bell, K.~L., Gilbody, H.~B., Hughes, J.~G., Kingston, A.~E. \& Smith, F.J.\ 1983,
Journal of Physical and Chemical Reference Data, 12:891-916,
DOI: 10.1063/1.555700. 

\bibitem[Bhardwaj(2003)]{bhardwaj2003}
{Bhardwaj}, A.\ 2003,
\grl, 30:2244, 
DOI: 10.1029/2003GL018495.

\bibitem[Ben-Jaffel et al.(2022)]{ben-jaffeletal2022}
{Ben-Jaffel}, L., 
{Ballester}, G.~E., 
{Garc{\'\i}a Mu{\~n}oz}, A., 
{Lavvas}, P.,
{Sing}, D.~K., et al.\ 2022,
Nature Astronomy, 6:141-153, 
DOI: 10.1038/s41550-021-01505-x.

\bibitem[Benneke et al.(2019)]{bennekeetal2019}
{Benneke}, B., {Wong}, I., {Piaulet}, C., {Knutson}, H.~A., {Lothringer}, J.\ 2019,
\apjl, 887:L14, 
DOI: 10.3847/2041-8213/ab59dc.

\bibitem[Bitsch et al.(2021)]{bitschetal2021}
{Bitsch}, B., {Raymond}, S.~N., {Buchhave}, L.~A., {Bello-Arufe}, A., {Rathcke}, A.~D. \& {Schneider}, A.~D.\ 2021,
\aap, 649:L5,
DOI: 10.1051/0004-6361/202140793.

\bibitem[Burkholder et al.(2019)]{burkholderetal2019}
Burkholder, J.~B., 
Sander, S.~P., 
Abbatt, J., 
B.~J.~R., et al.\ 2019,
Chemical Kinetics and Photochemical Data for Use
in Atmospheric Studies, Evaluation No. 19, JPL
Publication 19-5, Jet Propulsion Laboratory, Pasadena.

\bibitem[Carleo et al.(2021)]{carleoetal2021}
{Carleo}, I., {Youngblood}, A., {Redfield}, S., 
{Casasayas Barris}, N., {Ayres}, T.~R., et al.\ 2021,
\aj, 161:136, 
DOI: 10.3847/1538-3881/abdb2f.

\bibitem[Cecchi-Pestellini et al.(2006)]{cecchi-pestellinietal2006}
{Cecchi-Pestellini}, C., {Ciaravella}, A. \& {Micela}, G.\ 2006,
\aap, 458:L13-L16,
DOI: 10.1051/0004-6361:20066093.

\bibitem[Cecchi-Pestellini et al.(2009)]{cecchi-pestellinietal2009}
{Cecchi-Pestellini}, C., {Ciaravella}, A., {Micela}, G., {Penz}, T., A.~D. \& {Schneider}, A.~D.\ 2009,
\aap, 496:863-868,
DOI: 10.1051/0004-6361/200809955.

\bibitem[Chassefi{\`e}re(1996)]{chassefiere1996}
{Chassefi{\`e}re}, E.\ 1996, 
\jgr, 101, E11:26039-26056, 
DOI: 10.1029/96JE01951.

\bibitem[Dalgarno \& Lejeune(1971)]{dalgarnolejeune1971}
Dalgarno, A. \& Lejeune, G.\ 1971,
Planet. Space Sci., 19:1653:1667,
DOI: 10.1016/0032-0633(71)90126-7.

\bibitem[Dalgarno et al.(1999)]{dalgarnoetal1999}
{Dalgarno}, A., {Yan}, M. \& {Liu}, W.\ 1999,
\apj, 125:237-256, 
DOI: 10.1086/313267.

\bibitem[Delrez et al.(2021)]{delrezetal2021}
{Delrez}, L., {Ehrenreich}, D., {Alibert}, Y., {Bonfanti}, A., {Borsato}, L., {Fossati}, L., et al.\ 2021,
Nature Astronomy, 5:775-787,
DOI: 10.1038/s41550-021-01381-5.

\bibitem[Evans et al.(2017)]{evansetal2017}
{Evans}, T.~M., {Sing}, D.~K., {Kataria}, T., {Goyal}, J., {Nikolov}, N., et al.\ 2017,
Nature, 548:58-61,
DOI: 10.1038/nature23266.

\bibitem[Gandolfi et al.(2018)]{gandolfietal2018}
{Gandolfi}, D., {Barrag{\'a}n}, O., {Livingston}, J.~H., {Fridlund}, M., {Justesen}, A.~B., {Redfield}, S., et al.\ 2018,
\aap, 619:L10, DOI: 10.1051/0004-6361/201834289.

\bibitem[Garc\'ia Mu\~noz \& Schneider(2019)]{garciamunozschneider2019} 
Garc\'ia Mu\~noz, A. \&
Schneider, P.C.\ 2019, 
\apjl, 884:L43, 
DOI: 10.3847/2041-8213/ab498d.

\bibitem[Garc\'ia Mu\~noz et al.(2020)]{garciamunozetal2020} 
{Garc{\'\i}a Mu{\~n}oz}, A., {Youngblood}, A., {Fossati}, L., {Gandolfi}, D., {Cabrera}, J. \& {Rauer}, H.\ 2020,
\apjl, 888:L21,
DOI: 10.3847/2041-8213/ab61ff.

\bibitem[Garc\'ia Mu\~noz et al.(2021)]{garciamunozetal2021} 
{Garc{\'\i}a Mu{\~n}oz}, A., {Fossati}, L., {Youngblood}, A., {Nettelmann}, N., {Gandolfi}, D., {Cabrera}, J. \& {Rauer}, H.\ 2021,
\apjl, 907:L36, 
DOI: 10.3847/2041-8213/abd9b8.

\bibitem[Garc\'ia Mu\~noz(2023)]{garciamunoz2023} 
{Garc{\'\i}a Mu{\~n}oz}, A.\ 2023,
Icarus, 392:115373,
DOI: 10.1016/j.icarus.2022.115373.

\bibitem[Gillon et al.(2017)]{gillonetal2017}
{Gillon}, M., {Triaud}, A.~H.~M.~J., {Demory}, B.-O., {Jehin}, E., {Agol}, E. et al.\ 2017,
\nat, 542:456-460,
DOI: 10.1038/nature21360.

\bibitem[Guo \& Ben-Jaffel(2016)]{guoben-jaffel2016}
{Guo}, J.~H. \& {Ben-Jaffel}, L.\ 2016,
\apj, 818:107, 
DOI: 10.3847/0004-637X/818/2/107.

\bibitem[Guo(2019)]{guo2019}
{Guo}, J.~H.\ 2019,
\apj, 872:99,
DOI: 10.3847/1538-4357/aaffd4.

\bibitem[Green et al.(1977)]{greenetal1977}
Green, A.~E.~S., Jackman, C.~H. \& Garvey, R.~H.\ 1977,
\jgr, 82:5104--5111,
DOI: 10.1029/JA082i032p05104.

\bibitem[Huang et al.(2018)]{huangetal2018}
{Huang}, Ch. X., {Burt}, J., {Vanderburg}, A., {G{\"u}nther}, M.N., {Shporer}, A., {Dittmann}, J.A., et al.\ 2018,
\apjl, 868:L39,
DOI: 10.3847/2041-8213/aaef91.

\bibitem[Itikawa(1978)]{itikawa1978}
Itikawa, Y.\ 1978, 
Atomic Data and Nuclear Data Tables, 21:69-75, 
DOI: 10.1016/0092-640X(78)90004-9.

\bibitem[Itikawa et al.(1985)]{itikawaetal1985}
Itikawa, Y., Hara, S., Kato, T., Nakazaki, S., Pindzola, M.~S. \& Crandall, D.~H.\ 1985,
Atomic Data and Nuclear Data Tables, 33:149-193, 
DOI: 10.1016/0092-640X(85)90025-7.

\bibitem[Itikawa \& Ichimura(1990)]{itikawaichimura1990}
Itikawa, Y. \& Ichimura, A.\ 1990, 
Journal of Physical and Chemical Reference Data, 19:637-651,
DOI: 10.1063/1.555857.

\bibitem[Jamieson et al.(1992)]{jamiesonetal1992}
{Jamieson}, M.~J., {Finch}, M., {Friedman}, R.~S. \& {Dalgarno}, A.\ 1992,
\planss, 40:1719-1721, 
DOI: 10.1016/0032-0633(92)90128-B.

\bibitem[Johnstone(2020)]{johnstone2020}
{Johnstone}, C.~P.\ 2020,
\apj, 890:79,
DOI: 10.3847/1538-4357/ab6224.

\bibitem[Kimura \& Ikoma(2022)]{kimuraikoma2022}
{Kimura}, T. \& {Ikoma}, M.,\ 2022,
Nature Astronomy, 
DOI: 10.1038/s41550-022-01781-1.

\bibitem[Kozma \& Fransson(1992)]{kozmafransson1992}
Kozma, C. \& Fransson, C.\ 1992,
\apj, 390:602--621.

\bibitem[Krems et al.(2006)]{kremsetal2006}
{Krems}, R.~V., {Jamieson}, M.~J. \& {Dalgarno}, A.\ 2006,
\apj, 647:1531-1534,
DOI: 10.1086/504892.

\bibitem[Kurosaki et al.(2014)]{kurosakietal2014}
Kurosaki, K., Ikoma, M. \& Hori, Y.\ 2014,
\aap, 562:A80,
DOI: 10.1051/0004-6361/201322258.

\bibitem[Laher \& Gilmore(1990)]{lahergilmore1990}
Laher, R.~R. \& Gilmore, F.~R.\ 1990,
Journal of Physical and Chemical Reference Data, 19:277-305,
DOI: 10.1063/1.555872.

\bibitem[Lichtenegger et al.(2016)]{lichteneggeretal2016}
{Lichtenegger}, H.~I.~M., {Kislyakova}, K.~G., {Odert}, P., {Erkaev}, N.~V.,  {Lammer}, H., {Gr{\"o}ller}, H., {Johnstone}, C.~P., {Elkins-Tanton}, L., {Tu}, L.,  {G{\"u}del}, M. \& {Holmstr{\"o}m}, M.\ 2016,
\jgr (Space Physics), 121:4718-4732,
DOI: 10.1002/2015JA022226.

\bibitem[Locci et al.(2022)]{loccietal2022}
{Locci}, D., {Petralia}, A., {Micela}, G., {Maggio}, A.,  {Ciaravella}, A. \& {Cecchi-Pestellini}, C.\ 2022,
Planet. Science Journal, 3:1, 
DOI: 10.3847/PSJ/ac3f3c.

\bibitem[Luger \& Barnes(2015)]{lugerbarnes2015}
{Luger}, R. and {Barnes}, R.\ 2015,
Astrobiology, 15:119-143, 
DOI: 10.1089/ast.2014.1231.

\bibitem[Luque \& Pall\'e(2022)]{luquepalle2022}
{Luque}, R. \& {Pall{\'e}}, E.\ 2022,
Science, 377:1211-1214,
DOI: 10.1126/science.abl7164.

\bibitem[Madhusudhan et al.(2020)]{madhusudhanetal2020}
{Madhusudhan}, N., {Nixon}, M.~C., {Welbanks}, L., {Piette}, A.~A.~A. \& {Booth}, R.~A.\ 2020,
\apjl, 891:L7,
DOI: 10.3847/2041-8213/ab7229.

\bibitem[Mousis et al.(2020)]{mousisetal2020}
{Mousis}, O., {Deleuil}, M., {Aguichine}, A., {Marcq}, E., {Naar}, J., Acu{\~n}a Aguirre, L.\ 2020,
\apjl, 896:L22,
DOI: 10.3847/2041-8213/ab9530

\bibitem[Nakayama et al.(2022)]{nakayamaetal2022}
{Nakayama}, A. \& {Ikoma}, M. \& {Terada}, N.\ 2022,
\apj, 937:72, 
DOI: 10.3847/1538-4357/ac86ca

\bibitem[Nettelmann et al.(2010)]{nettelmannetal2010}
{Nettelmann}, N., {Kramm}, U., {Redmer}, R. \& {Neuh{\"a}user}, R.\ 2010,
\aap, 523:A26,
DOI: 10.1051/0004-6361/200911985.

\bibitem[Rockcliffe et al.(2021)]{rockcliffeetal2021}
{Rockcliffe}, K.~E., 
{Newton}, E.~R., 
{Youngblood}, A., 
{Bourrier}, V.,
{Mann}, A.~W., 
{Berta-Thompson}, Z., 
{Ag{\"u}eros}, M.~A., 
{N{\'u}{\~n}ez}, A. \& {Charbonneau}, D.\ 2021,
\aj, 162:116, 
DOI: 10.3847/1538-3881/ac126f.

\bibitem[Shaikhislamov et al.(2020)]{shaikhislamovetal2020}
{Shaikhislamov}, I.~F., {Fossati}, L., {Khodachenko}, M.~L., {Lammer}, H., {Garc{\'\i}a Mu{\~n}oz}, A.,  {Youngblood}, A., {Dwivedi}, N.~K. \& {Rumenskikh}, M.~S.\ 2020,
{\aap}, 639:A109,
DOI: 10.1051/0004-6361/202038363.

\bibitem[Shematovich et al.(2014)]{shematovichetal2014}
{Shematovich}, V.~I. and {Ionov}, D.~E. and {Lammer}, H.\ 2014,
{\aap}, 571:A94,
DOI: 10.1051/0004-6361/201423573.

\bibitem[Slinker et al.(1988)]{slinkeretal1988}
Slinker, S.P., Taylor, R.D. \& Ali, A.W.\ 1988,
J. Appl. Phys., 63(1):1--10,
DOI: 10.1063/1.340491.

\bibitem[Tayal \& Zatsarinny(2016)]{tayalzatsarinny2016}
Tayal, S.~S. \& {Zatsarinny}, O.\ 2016,
Phys. Review A, 94:042707, 
DOI: 10.1103/PhysRevA.94.042707.

\bibitem[Tian et al.(2008)]{tianetal2008}
{Tian}, F., {Solomon}, S.~C., {Qian}, L., {Lei}, J. \&{Roble}, R.~G.\ 2008,
\jgr, 113:E07005, 
DOI: 10.1029/2007JE003043.

\bibitem[Tsiaras et al.(2019)]{tsiarasetal2019}
{Tsiaras}, A., {Waldmann}, I.~P., {Tinetti}, G., {Tennyson}, J. \& {Yurchenko}, S.~N.\ 2019,
Nature Astronomy, 3:1086-1091,
DOI: 10.1038/s41550-019-0878-9.

\bibitem[Victor et al.(1994)]{victoretal1994}
Victor, G.~A., Raymond, J.~C. \& Fox, J.~L.\ 1994,
Astrophys. J., 435:864--869,
DOI: 10.1086/174866.

\bibitem[Otegi et al.(2020)]{otegietal2020}
{Otegi}, J.~F., {Bouchy}, F. \& {Helled}, R.\ 2020,
\aap, 634:A43,
DOI: 10.1051/0004-6361/201936482.

\bibitem[Peterson et al.(1978)]{petersonetal1978}
{Peterson}, L.~R., {Garvey}, R.~H. \& {Green}, A.~E.~S.\ 1978,
\jgr, 83:5315-5318,
DOI: 10.1029/JA083iA11p05315

\bibitem[Tinetti et al.(2018)]{tinettietal2018}
{Tinetti}, G., {Drossart}, P., {Eccleston}, P., {Hartogh}, P.,  {Heske}, A., et al.\ 2018,
Experimental Astronomy, 46:135-209,
DOI: 10.1007/s10686-018-9598-x

\bibitem[Valencia et al.(2010)]{valenciaetal2010}
Valencia, D., Ikoma, M., Guillot, T. \& Nettelmann, N.\ 2010,
\aap, 516:A20,
DOI: 10.1051/0004-6361/200912839.

\bibitem[Venturini et al.(2020)]{venturinietal2020}
{Venturini}, J. \& {Guilera}, O.~M., {Haldemann}, J. \& {Ronco}, M.~P. \& {Mordasini}, Ch.\ 2020,
\aap, 643:L1,
DOI: 10.1051/0004-6361/202039141.

\bibitem[Wilson et al.(2021)]{wilsonetal2021}
{Wilson}, D.~J., {Froning}, C.~S., {Duvvuri}, G.~M., {France}, K., {Youngblood}, A., et al.\ 2021,
\apj, 911:18, 
DOI: 10.3847/1538-4357/abe771

\bibitem[Yoshida et al.(2022)]{yoshidaetal2022}
{Yoshida}, T., {Terada}, N., {Ikoma}, M. \& {Kuramoto}, K.\ 2022,
\apj, 934:137, 
DOI: 10.3847/1538-4357/ac7be7.

\bibitem[Zahnle et al.(1988)]{zahnleetal1988}
{{Zahnle}, K.~J., {Kasting}, J.~F. \& {Pollack}, J.~B.}\ 1988,
\icarus, 74:62-97,
DOI: 10.1016/0019-1035(88)90031-0.

\bibitem[Zhang et al.(2022)]{zhangetal2022}
{Zhang}, M., 
{Knutson}, H.~A., 
{Wang}, L., 
{Dai}, F., 
{dos Santos}, L.~A., et al.\ 2022,
\aj, 163:68, 
DOI: 10.3847/1538-3881/ac3f3b.

\end{thebibliography}
\end{document}